\documentclass[12pt]{article}
\usepackage{cite}
\usepackage{CJK}
\usepackage{amssymb}
\usepackage{amsmath}
\usepackage{enumerate}
\usepackage{graphicx}
\usepackage{amsfonts}
\usepackage{url}
\usepackage{bm}

\usepackage{pifont}
\usepackage{epsfig,subfigure,dsfont,amsthm,amsbsy,mathrsfs,amscd}

\DeclareMathOperator{\Span}{span}

\def\la{\langle}
\def\ra{\rangle}
\def\be{\begin{equation}}
\def\ee{\end{equation}}
\def\ba{\begin{array}}
\def\ea{\end{array}}
\input amssym.def

\newcommand\btd{\raise 2pt \hbox{$\hat\bigtriangledown$}\hskip 1.5pt}
\newcommand\bt{\raise 2pt \hbox{$\bigtriangledown$}\hskip 1.5pt}
\begin{document}
 \title{\large\bf Non-commutativity and Local Indistinguishability of Quantum States}
\author{Teng MA$^\dag$, Ming-Jing ZHAO$^\ddag$, Yao-Kun WANG$^\diamond$$^\odot$ \&
 Shao-Ming FEI$^*$$^\circledast$$^\dag$\\[10pt]
\footnotesize
\small $^\dag$School of Mathematical Sciences, Capital Normal University,\\
\small Beijing 100048, P. R. China\\
\footnotesize
\small $^\ddag$Department of Mathematics, School of Science,\\
\small Beijing Information Science and Technology University, Beijing 100192, P. R. China\\
\small $^\diamond$Institute of Physics, Chinese Academy of Sciences, Beijing  100190, P. R. China\\
\small $^\odot$College of Mathematics, Tonghua Normal University,\\
\small Tonghua 134001, P. R. China\\
\small $^\circledast$Max-Planck-Institute for Mathematics in the Sciences,\\
\small Leipzig 04103, Germany}
\date{}

\maketitle

\centerline{$^\ast$ Correspondence to feishm@cnu.edu.cn}
\bigskip

\begin{abstract}

We study the local indistinguishability problem of quantum states.
By introducing an easily calculated quantity, non-commutativity, we
present an  criterion which is both necessary and
sufficient for the local indistinguishability of a complete set of
pure orthogonal product states. A constructive distinguishing procedure to obtain the
concrete local measurements and classical communications is given.
The non-commutativity of ensembles
can be also used to characterize the quantumness  for
classical-quantum or quantum-classical correlated states.
\end{abstract}
\bigskip

Nonlocality is one of the most important
features in quantum mechanics.  Quantum entanglement was firstly
introduced to characterize and quantify the nonlocality,  as it acts
as a crucial resource in many quantum information processing tasks
\cite{nielsen}. On the other hand, there are also quantum
information and computation processing tasks without quantum entanglement,
like the quantum computation with one qubit (DQC1) \cite{knil}.
Other types of quantum correlations such as quantum discord
\cite{ollivier} are introduced to capture this kind of quantum
advantage. The quantum discord is considered  more general than
entanglement and  captures the quantum correlations that
entanglement fails  to capture \cite{modi}.

A well known quantum phenomenon that exhibits  quantum nonlocality
is the local  indistinguishability for some quantum states. State
discrimination or distinguishing is essentially primitive for many
quantum information tasks, such as quantum cryptography
\cite{bennett3} and quantum algorithms \cite{bergou}. Moreover,
with the remarkable experimental advances in preparation and
measurement of quantum states \cite{raimond,haroche}, it becomes
essential to have a theory to assess the performance of quantum
state discrimination protocols. One basic problem of state
discrimination is  judging the local distinguishability for a set of
pure orthogonal product states (POPS). Consider a set of bipartite
POPS, each with a prior probability, which constitutes an ensemble
of a density operator $\rho_{AB}$. Although a set of POPS can always
be distinguished globally, it may not be distinguished
locally by local operations and classical communications (LOCC). This is
called \emph{quantum nonlocality without entanglement}
\cite{bennett2}. In \cite{brod} the relation between quantum discord
and local indistinguishability has been investigated. However,
contrary to one's intuitive appeal, the quantum discord of
$\rho_{AB}$ is not an indicator of local indistinguishability for a
set of states that constitute the pure-state decomposition of
$\rho_{AB}$ \cite{brod,modi}. It shows no relation between zero
quantum correlation and the local distinguishability for a set of
POPS. The local indistinguishability problem of a set of POPS
remains open, although many research have been done so far
\cite{bennett,bennett2,walgate1,walgate2,ghosh,chen,brod}.

It is natural to ask what kind of quantity or quantumness accounts
for the quantum nonlocality. In this article, instead of quantum
correlation measures, such as quantum discord, quantum deficit,
etc., we investigate local indistinguishability for a set of POPS
from the point of quantumness of ensemble. We first introduce an
easily calculated quantity, non-commutativity, to quantify the
quantumness of a quantum ensemble. Based on the non-commutativity,
we present a necessary and sufficient criterion for the local
indistinguishability. We also give a constructive distinguishing procedure to judge
the local indistinguishability for any given set of POPS by using
the criterion. Moreover, by proving the uniqueness of the expression of
for semi-classical quantum correlated states, we show that our
definition for quantumness of ensembles can be used to characterize
the quantumness for semi-classical states.

\medskip
\noindent{\bf Results}

\medskip
{\sf Non-commutativity for a set quantum states}~
A quantum ensemble containing only two
pure states $|\psi\ra$ and $|\phi\ra$ with equal probability
can be viewed as a set of binary signals in some communication
scheme. If $|\psi\ra$ and $|\phi\ra$ are orthogonal, the ensemble becomes
classical since different states of  classical information can be
thought of merely as orthogonal quantum states \cite{nielsen}. Non
complete overlap of the two states, $x=|\la \psi|\phi \ra |$, $0<x<1$, induces
quantumness for the ensemble. In fact the ensemble is most
`quantum' when $x=1/\sqrt{2}$, i.e. when the two states have an angle $45^{\circ}$ between them \cite{fuchs}.

Taking the above observation into account, we introduce
non-commutativity to characterize the quantumness for a quantum ensemble. Let
${A_1,A_2,...,A_n}$ be a set of operators. We define the total non-commutativity for this set,
\begin{equation}\label{nc}
N(A_1,A_2,...,A_n)=\sum\limits_{i,j=1,i>j}^n||[A_i,A_j]||,
\end{equation}
where $[A,B]=AB-BA$, $||A||$ is the trace norm of the operator $A$, $||A||=Tr\sqrt{AA^\dagger}$.
The non-commutativity is a natural measure of `quantumness' for a quantum
ensemble $\varepsilon=\{p_i, \rho_i\}$, where $\rho_i$ are density
operators with probability $p_i$. Denote $A_i=p_i\rho_i$, then
$N(A_1,A_2,...,A_n)$ is the measure of quantumness for the ensemble $\varepsilon$.
Here for the problem of local distinguishability of states $\rho_1,\rho_2,...,\rho_n$,
the prior probabilities $p_i$ are irrelevant. One only needs to concern the quantity
$N(\varepsilon)=N(\rho_1,\rho_2,...,\rho_n)$ to judge the local distinguishability of the set of states $\rho_1,\rho_2,...,\rho_n$.
The prior probabilities $p_i$ do make sense when one concerns the quantumness of a state
given by the ensemble $\varepsilon=\{p_i, \rho_i\}$ (see section ``The quantumness of semi-classical states").

The non-commutativity $N$ has the following properties, which make
it a well defined measure for quantumness of an ensemble:

(1) $N$ is non negative;

(2) $N$ is unitary invariant, $N(\varepsilon)=N(U\varepsilon U^{\dagger})$, where $U\varepsilon U^{\dagger}=\{p_i, U\rho_i U^{\dagger}\}$ with $U$ being any unitary matrix;

(3) For an  ensemble  only containing two pure states $A_1=|\psi\ra\la\psi|$ and
$A_2=|\phi\ra\la\phi|$, without considering their prior probabilities, $N(A_1,A_2)$ is zero only when the overlap
$x=|\la \phi|\psi\ra|=0$ or $1$. $N(A_1,A_2)$ gets to a maximum of
$1$ when $x={1}/{\sqrt{2}}$ (see Methods for the
proof), which coincides with the above analysis of quantumness for
two pure states;

(4) The sum of two sets' non-commutativity is equal
to or less than the non-commutativity of the sum of the two sets:
$N(\{|a_i\ra\})+N(\{|b_i\ra\})\leq N(\{\{|a_i\ra\},\{|b_i\ra\}\})$.
One can easily verify the inequality from the definition of $N$. The equality holds if
$\{|a_i\ra\}$ and $\{|b_i\ra\}$ are either mutually orthogonal or
identical.

A set of quantum states
corresponds to a quantum ensemble. Nevertheless, one density
operator may have many quantum ensemble decompositions. For
instance, consider the density operator in a $3\otimes3$ system,
$\rho=\frac{1}{9}\mathbb{I}=\frac{1}{9}\sum\limits_{i=1}^9|\psi_i\ra
\la\psi_i|=\frac{1}{9}\sum\limits_{s,t=1}^3|st\ra\la\ st|$, where
$|\psi_1\ra=|1\ra|1\ra,
|\psi_{2,3}\ra=\frac{1}{\sqrt{2}}|0\ra|0\pm1\ra,
|\psi_{4,5}\ra=\frac{1}{\sqrt{2}}|2\ra|1\pm2\ra,
|\psi_{6,7}\ra=\frac{1}{\sqrt{2}}|1\pm2\ra|0\ra,
|\psi_{8,9}\ra=\frac{1}{\sqrt{2}}|0\pm1\ra|2\ra$,
$|st\ra=|s\ra\otimes|t\ra$, $|s\ra$ and $|t\ra$ are the
computational basis. Both $\{|\psi_i\ra\}$ and $\{|st\ra\}$ are
pure-state decompositions of $\rho$. However, the set $\{|st\ra\}$
can be locally distinguished while $\{|\psi_i\ra\}$ cannot
\cite{bennett2}. Hence that $\rho$ having zero quantum correlation
is not an indicator for the local distinguishability of the states
 in $\rho$'s pure-state decomposition. On the other hand, if we change the probability of
$|\psi_i\ra$, $\rho$ is no longer an identity and can have nonzero
quantum correlation, but the nine states $|\psi_i\ra$ still remain
locally indistinguishable. Usually, the local indistinguishability
of a set of pure states has no simple relations with the properties
of the related density operator \cite{modi}.

To study the local indistinguishability, and its relations with the quantumness of a quantum
ensemble in terms of non-commutativity, in the following we call an ensemble $\varepsilon=\{p_i,
|\psi_i\ra\}$ classical if $N(\varepsilon)=0$, and quantum if
$N(\varepsilon)>0$. If $N(\varepsilon)=0$, from the properties of
non-commutativity, $\{|\psi_i\ra\}$ must be either mutually
orthogonal or identical and the states form a set of classical
signals from informatics point of view. If $N(\varepsilon)>0$, among
$\{|\psi_i\ra\}$ there must be at least one pair of  states that are
neither orthogonal nor identical. Consider a set of bipartite POPS
$\{|\psi_i\ra=|a_i\ra\otimes|b_i\ra\}$, where $|\psi_i\ra$ are all
mutually orthogonal, $|a_i\ra$ and $|b_i\ra$ are associated to
partite $A$ and $B$ respectively. Obviously, this set forms a
classical ensemble $\varepsilon=\{p_i, |\psi_i\ra \}$ since
$N(\varepsilon)=0$, where $p_i$ is the probability with respect to
$|\psi_i\ra$. Correspondingly one has ensembles
$\varepsilon^A=\{p_i, |a_i\ra \}$ and $\varepsilon^B=\{p_i, |b_i\ra
\}$ respectively. If $N(\varepsilon^A)=N(\varepsilon^B)=0$, we call
$\varepsilon$ a classical-classical ensemble. If
$N(\varepsilon^A)=0$, $N(\varepsilon^B)>0$, we call $\varepsilon$
a classical-quantum ensemble. Analogously we can define quantum-classical and
quantum-quantum ensembles.

\medskip
{\sf Local distinguishability for a set quantum states}~
A set of states is said to be reliably distinguished locally or distinguished locally
if all the states in the set can be distinguished by \textit{finite} rounds of LOCC protocols. If there
are at least two states in a set which can not be distinguished by finite rounds of LOCC protocols, we say that the
set can not be distinguished locally. The arguably more operational ``asymptotic local operations and classical
communications discrimination problem" \cite{kleinmann,childs} is not under our consideration.

{\bf Theorem 1}~If a set of bipartite pure orthogonal product states cannot be
locally distinguished, the ensemble composed of these states must be
a quantum-quantum ensemble.

\noindent{\sf Proof.}~ Consider a set of bipartite POPS, not necessarily complete,
$\{|\psi_i \ra =|a_i \ra \otimes|b_i\ra\}$ with $\la
\psi_i|\psi_j\ra=0,~\forall{i\neq j}$. Suppose these states form a
classical-quantum or classical-classical ensemble
$\varepsilon=\{p_i, |a_i\ra\otimes |b_i\ra\}$ with non zero $p_i$.
Since $N(\{|a_i\ra\})=0$, $\{|a_i\ra\}$ must be either mutually
orthogonal or identical. To distinguish $\{|\psi_i\ra\}$ locally, we
can first take  projective measurement on $A$ side to distinguish
those orthogonal states in $\{|a_i\ra\}$. For those identical states
$|a_i\ra$, for example, $|a_1\ra=|a_2\ra$, we can take a projective
measurement on $B$ to distinguish state $|a_1\ra$ from state
$|a_2\ra$, since $|b_1\ra$ and $|b_2\ra$ must be mutually orthogonal
to ensure that $\la \psi_1|\psi_2\ra=0$. In this way we can
distinguish the states in the set $\varepsilon$ reliably. The
analysis is similar if $\varepsilon$ is a quantum-classical
ensemble. Therefore if a set of bipartite POPS forms a
classical-classical, quantum-classical or classical-quantum
ensemble, the set can be locally distinguished. And if the set of
POPS states cannot be locally distinguished, the ensemble composed
of these states must be a quantum-quantum one. \quad $\Box$

Since it is impossible to form a quantum-quantum ensemble for a set
of POPS in a $2\otimes2$ system, Theorem 1 is both necessary and
sufficient for $2\otimes 2$ systems. For higher dimensional cases,
there are quantum-quantum ensembles whose states can be locally
distinguished. In fact, for $2\otimes n$ systems, any set of POPS is
locally distinguishable \cite{bennett}. Theorem 1 reveals a relation
between local distinguishability and quantumness of a set of states.
In the following, we give a necessary and sufficient criterion for
local distinguishability for all POPS from the view of quantumness
of ensembles.

We say that two sets $\{|{a}_i\ra\}$ and $\{|{a'}_j\ra\}$ are
orthogonal if and only if $\la {a}_i|{a'}_j \ra=0$,
$\forall\,{i,j}$. If  a set of states cannot be divided into subsets
such that those subsets are mutually orthogonal, then we call the
set a single set. Consider a set of states
$\varepsilon=\{|a_1\ra=|0\ra$, $|a_2\ra=|0+1\ra$, $|a_3\ra=|2\ra\}$.
This set can be divided into two single subsets
$\varepsilon_1=\{|a_1\ra, |a_2\ra\}$ and $\varepsilon_2=\{|a_3\ra\}$
which are orthogonal and  each subset cannot be split further. We
call this partition the direct sum decomposition and denote it as
$\varepsilon=\varepsilon_1\oplus \varepsilon_2$. A single set means
that the set cannot be decomposed into such direct sums.
If all the subsets are single sets in a direct sum decomposition,
we say that the decomposition is a single set decomposition. It can verified that
the single set decomposition of a non-single set is unique. For a
single set, adding some states in the vector space spanned by itself
will keep the set a single one. Nevertheless, taking away some
states from a single set, the set could become a non-single one.
In the following, by a set of states' decomposition we mean the single set decomposition.

{\bf Lemma} For a set of states $\varepsilon=\{|a_i\ra\}$, $m=\dim (\Span
\{|a_i\ra\})$, the following statements are equivalent: (a)
$\varepsilon$ is a single set. (b) There are $m$ linear independent
states $\{|a_{i_k}\ra\}_{k=1}^{m}$ in $\varepsilon$ satisfying the
following relations:
\begin{equation}\label{nn}
\begin{aligned}
0&<N(|a_{i_1}\ra, |a_{i_2}\ra)<N(|a_{i_1}\ra, |a_{i_2}\ra,|a_{i_3}\ra)\\
&<...<N(|a_{i_1}\ra,|a_{i_2}\ra,...,|a_{i_m}\ra).
\end{aligned}
\end{equation}
(c) A nondestructive projective  measurement, a measurement which keeps the quantum state unchanged \cite{chen}, can do nothing to
distinguish the states in $\varepsilon$.

See Methods for the proof of the Lemma. From the Lemma we have

{\bf Theorem 2}~{For a complete set of $m\otimes n$ POPS, $\varepsilon=\{|\psi_i \ra =|a_i
\ra \otimes|b_i\ra\}$ with $\la \psi_i|\psi_j\ra=0,~\forall{i\neq
j}$, the set $\varepsilon$
 cannot be completely  locally distinguished if and only if there exist  subsets
$\{|\psi'_i\ra=|a'_i\ra\otimes |b'_i\ra\}\subseteq\varepsilon$, such
that  $\{|a'_i\ra\}$ and $\{|b'_i\ra\}$ are all single sets, i.e.,
there exist $m'=\dim(\Span\{|a'_i\ra\})$ linear independent
$\{|a_{i_k}\ra\}_{k=1}^{m'}$ in $\{|a'_i\ra\}$ and
$n'=\dim(\Span\{|b'_i\ra\})$ linear independent
$\{|b_{j_l}\ra\}_{l=1}^{n'}$ in $\{|b'_i\ra\}$ satisfying
\begin{equation}\label{nn2}
\begin{aligned}
0&<N(|a_{i_1}\ra, |a_{i_2}\ra)<N(|a_{i_1}\ra, |a_{i_2}\ra,|a_{i_3}\ra)\\
&<...<N(|a_{i_1}\ra,|a_{i_2}\ra,...,|a_{i_{m'}}\ra),\\[2mm]
0&<N(|b_{j_1}\ra, |b_{j_2}\ra)<N(|b_{j_1}\ra, |b_{j_2}\ra, |b_{j_3}\ra)\\
&<...<N(|b_{j_1}\ra, |b_{j_2}\ra,..., |b_{j_{n'}}\ra).
\end{aligned}
\end{equation}}

\noindent {\sf Proof.}~ Recall that a complete set of POPS can be locally distinguished if
and only if the states can be distinguished by local nondestructive
projective measurement and classical communication \cite{chen}.  If
a set of states in $\varepsilon$ cannot be completely locally
distinguished by local nondestructive projective measurement and
classical communication, there must exist a subset
$\{|\psi'_i\ra=|a'_i\ra\otimes |b'_i\ra\}$ of $\varepsilon$ such
that $\{|a'_i\ra\}$ and $\{|b'_i\ra\}$ are all single sets (Lemma 1).
Obviously, the converse is also right.  From Lemma 1 we have that
the formula (\ref{nn2}) holds if and only if $\{|a'_i\ra\}$ and
$\{|b'_i\ra\}$ are single sets. \quad $\Box$

From the view of accessible information, the more quantum  an
ensemble is, the less information one can get from the ensemble
\cite{nielsen,fuchs}. Hence the quantumness of the two parts of a
complete set of POPS must be `` large " enough, at least larger than
$N(\{|a_{i_k}\ra\})$ and $N(\{|b_{j_l}\ra\})$, so that one can only
get limited information and the states in the set cannot be locally
distinguished. Therefore the local quantumness of an ensemble
determines the local indistinguishability.

Theorem 2 gives a constructive distinguishing procedure to judge the local
distinguishability for a complete set of POPS.  Let $\{|\psi_i\ra=|a_i\ra
\otimes|b_i\ra\}$ be a  complete set of POPS, corresponding to two
sets $\varepsilon^A=\{|a_i\ra\}$ and $\varepsilon^B=\{|b_i\ra\}$. To
verify the inequality (\ref{nn2}), one needs to find all the subsets
involved. This can be done in the following way.

(i) Decompose the sets $\varepsilon^A$ and $\varepsilon^B$
into subsets, $\varepsilon^A=\varepsilon^A_1\oplus
\varepsilon^A_2\oplus...$, $\varepsilon^B=\varepsilon^B_1\oplus
\varepsilon^B_2\oplus...$ with the states in each subset
constituting a single set. This process employs the corresponding
local measurement $M^A=1P^A_1\oplus 2P^A_2\oplus...$ and
$M^B=1P^B_1\oplus 2P^B_2\oplus...$, where
$P^A_1,P^A_2,P^B_1,P^B_2,...$ is the projections to the spaces
spanned by
$\varepsilon^A_1,\varepsilon^A_2,\varepsilon^B_1,\varepsilon^B_2,...$
respectively.

(ii) Find the overlapped states between these subsets,
$\{|\psi_s\ra\}_{ij}=\varepsilon^A_i\widetilde{\cap}
\varepsilon^B_j$, where $\widetilde{\cap}$ means to find the states
with the same subscripts by classical communication. For example, if
$\varepsilon^A_1=\{|a_1\ra, |a_2\ra, |a_3\ra\}$,
$\varepsilon^B_1=\{|b_2\ra, |b_3\ra, |b_4\ra\}$, then
$\varepsilon^A_1\widetilde{\cap} \varepsilon^B_1=\{|\psi_2\ra,
|\psi_3\ra\}_{11}$. Each $\{|\psi_s\ra=|a_s\ra\otimes
|b_s\ra\}_{ij}$ corresponds to two new subsets
$\varepsilon^{A_{ij}}=\{|a_s\ra\}$ and
$\varepsilon^{B_{ij}}=\{|b_s\ra\}$.

We repeat the above process $n$
rounds for the new subsets until each of those new subsets has only
one element or both $A$ and $B$ parts cannot be decomposed further.
At last we have  subset
$\{|\psi_k\ra=|a_k\ra\otimes|b_k\ra\}_{st,...,ij}$ (here $st,...,ij$
are the measurement outcomes) corresponding to two new sets
$\varepsilon^{A_{st,...,ij}}=\{|a_k\ra\}$ and
$\varepsilon^{B_{st,...,ij}}=\{|b_k\ra\}$, where both
$\varepsilon^{A_{st,...,ij}}$ and $\varepsilon^{B_{st,...,ij}}$ are  single sets. If all the final single subsets
have one element, the set $\{|\psi_i\ra\}$ can be locally
distinguished.

As an example, there is a complete set of POPS for $3\otimes4$
system,
\begin{equation}\label{set34}
\begin{aligned}&|\psi_1\ra=|1\ra|1\ra,   \hspace{2.1cm} |\psi_{8,9}\ra=|0\pm1\ra|2\ra/\sqrt{2}, \\
&|\psi_{2,3}\ra=|0\ra|0\pm1\ra/\sqrt{2},          \hspace{0.7cm} |\psi_{10}\ra=|0\ra|3\ra,  \\
&|\psi_{4,5}\ra=|2\ra|1\pm2\ra/\sqrt{2},          \hspace{0.7cm} |\psi_{11}\ra=|1\ra|3\ra,\\
&|\psi_{6,7}\ra=|1\pm2\ra|0\ra/\sqrt{2},
  \hspace{0.7cm}|\psi_{12}\ra=|2\ra|3\ra,
\end{aligned}
\end{equation}
which corresponds to two sets,
$\varepsilon^A=\{|a_1\ra=|1\ra,|a_{2,3}\ra=|0\ra,|a_{4,5}\ra=|2\ra,|a_{6}\ra=\frac{1}{\sqrt{2}}|1+2\ra
,|a_7\ra=\frac{1}{\sqrt{2}}|1-2\ra,|a_{8}\ra=\frac{1}{\sqrt{2}}|0+1\ra,|a_9\ra=\frac{1}{\sqrt{2}}|0-1\ra,
|a_{10}\ra=|0\ra,|a_{11}\ra=|1\ra,|a_{12}\ra=|2\ra\}$ and
$\varepsilon^B=\{|b_1\ra=|1\ra,|b_2\ra=\frac{1}{\sqrt{2}}|0+1\ra,|b_3\ra=\frac{1}{\sqrt{2}}|0-1\ra,|b_4\ra
=\frac{1}{\sqrt{2}}|1+2\ra,|b_5\ra=\frac{1}{\sqrt{2}}|1-2\ra,|b_{6,7}\ra=|0\ra,|b_{8,9}\ra=|2\ra,|b_{10,11,12}\ra=|3\ra\}$.
The distinguishing process is as follows.

Round 1: (i) There are
$3=\dim (\Span \varepsilon^A)$ linear independent states in
$\varepsilon^A$ satisfying $0<N(|a_1\ra, |a_6\ra)=1<N(|a_1\ra,
|a_6\ra, |a_8\ra)\doteq2.87$. So $\varepsilon^A$ is a single set.
Hence the first measurement should not be applied to $A$ side. We
denote $\varepsilon^A=\varepsilon^A_0$ (here 0 means to do no
measurement). For $B$ side, we have decomposition
$\varepsilon^B=\varepsilon^B_1\oplus \varepsilon^B_2$, where
$\varepsilon^B_1=\{|b_1\ra,...,|b_9\ra\}$ and
$\varepsilon^B_2=\{|b_{10}\ra,|b_{11}\ra,|b_{12}\ra\}$ are all
single sets. Hence $\varepsilon^B_1$ is a single set as there are
$3=\dim(\Span \varepsilon^B_1)$ linear independent states satisfying
$0<N(|b_1\ra, |b_2\ra)=1<N(|b_1\ra, |b_2\ra, |b_4\ra)\doteq2.87$.
This decomposition employs measurement $M^B=1(|0\ra \la 0|+ |1\ra
\la 1|+ |2\ra \la 2|)+2 |3\ra \la 3|$ to distinguish
$\varepsilon^B_1$ from $\varepsilon^B_2$.

(ii) Find the overlapped
states by classical communication, $\varepsilon^A_0\widetilde{\cap}\varepsilon^B_1=\{|\psi_1\ra,...,|\psi_9\ra\}_{01}$,
$\varepsilon^A_0\widetilde{\cap}\varepsilon^B_2=\{|\psi_{10}\ra,
|\psi_{11}\ra, |\psi_{12}\ra\}_{02}$. They correspond to four new
sets $\varepsilon^{A_{01}}=\{|a_1\ra,...,|a_9\ra\}$,
$\varepsilon^{B_{01}}=\{|b_1\ra,...,|b_9\ra\}$,
$\varepsilon^{A_{02}}=\{|a_{10}\ra, |a_{11}\ra, |a_{12}\ra\}$,
$\varepsilon^{B_{02}}=\{|b_{10}\ra, |b_{11}\ra, |b_{12}\ra\}$.

Round 2: (i) Do decomposition for  the above four new sets.
 Note $\varepsilon^{A_{01}}, \varepsilon^{B_{01}}$ are all single sets (the states $|a_4\ra, |a_6\ra, |a_8\ra$
 in $\varepsilon^{A_2,01}$ are linear independent and satisfy $0<N(|a_4\ra,|a_6\ra)=1<N(|a_4\ra, |a_6\ra, |a_8\ra)\doteq1.87$,
 and $\varepsilon^{B_{01}}=\varepsilon^{B}_1$). Therefore the set
 $\{|\psi_1\ra,...,|\psi_9\ra\}_{01}$ is the subset described in Theorem 2.
In fact they are the nine locally indistinguishable states in
\cite{bennett2}. But we can distinguish the set $\{|\psi_{10}\ra,
|\psi_{11}\ra, |\psi_{12}\ra\}_{02}$ further since we have
decomposition $\varepsilon^{A_{02}}=\varepsilon^{A_{02}}_1\oplus
\varepsilon^{A_{02}}_2\oplus \varepsilon^{A_{02}}_3$, where
$\varepsilon^{A_{02}}_1=\{|a_{10}\ra\}$,
$\varepsilon^{A_{02}}_2=\{|a_{11}\ra\}$,
$\varepsilon^{A_{02}}_3=\{|a_{12}\ra\}$. This employs measurement
$M^A=1|0\ra \la 0|+2|1\ra \la 1|+3|2\ra \la 2|$. However,
$\varepsilon^{B_{02}}$ is single, denote
$\varepsilon^{B_{02}}=\varepsilon^{B_{02}}_0$.

(ii) Find overlapped
states by classical communication,
$\varepsilon^{A_{02}}_1\widetilde{\cap}\varepsilon^{B_{02}}_0=\{|\psi_{10}\ra\}_{10,02}$,
$\varepsilon^{A_{02}}_2\widetilde{\cap}\varepsilon^{B_{02}}_0=\{|\psi_{11}\ra\}_{20,02}$,
$\varepsilon^{A_{02}}_3\widetilde{\cap}\varepsilon^{B_{02}}_0=\{|\psi_{12}\ra\}_{30,02}$.
All the above new  subsets either have one element each or both
sides are single ones.

Finally we can divide the set (\ref{set34})
into four parts, $\{|\psi_1\ra,...,|\psi_9\ra\}_{01}$,
$\{|\psi_{10}\ra\}_{10,02}$, $\{|\psi_{11}\ra\}_{20,02}$,
$\{|\psi_{12}\ra\}_{30,02}$, and the set cannot be completely
distinguished due to the subset
$\{|\psi_1\ra,...,|\psi_9\ra\}_{01}$.

\medskip
{\sf The quantumness of semi-classical states}~
The non-commutativity defined by
(\ref{nc}) can be also used to describe the quantumness for
bipartite semi-classical states. For a given semi-classical state,
$\rho_{qc}=\sum_i p_i \rho_i \otimes |i\ra \la i|$,
the corresponding ensemble $\varepsilon=\{p_i, \rho_i\}$ is fixed
(see Methods). Denote $X_i=p_i \rho_i$. We have the
non-commutativity describing the quantumness for the state
$\rho_{qc}$,
\begin{equation}\label{nqc}
N(\rho_{qc})=\sum_{i>j}||[X_i,X_j]||.
\end{equation}
It is easy to check that $N(\rho_{qc})$ satisfies the following
properties, which actually makes it a candidate for
quantum-correlation measures for semi-classical states \cite{modi}:
(a) it is positive, (b) it is zero for classical-classical
correlated states, (c) it is invariance under local-unitary
transformations, and (d) it is non-increasing when an ancillary
system is introduced. This quantity $N(\rho_{qc})$ describes the
quantumness of semi-classical states. It is different from those
quantum correlation measures, such as quantum discord, quantum
deficit, etc., which are based on the measurements and their
estimations involve extremely complicated optimization process.
Interestingly, while the quantity $N(\rho_{qc})$ can be easily computed,
it shows similar behavior to other quantum correlation measures for
semi-classical states. Hence,  instead of those quantum correlation
measures, one can use the non-commutativity to characterize the
quantum correlations of semi-classical states $\rho_{qc}$. Moreover,
non-commutativity provides a tool to explore the relation between
the ensemble quantumness and those quantum correlation measures.
Fig.1 shows the comparison between quantum correlations and ensemble
quantumness (non-commutativity) for a $3\otimes3$ system for which
it is extremely difficult to calculate quantum correlations
analytically.
\begin{figure}
  \centering
  \label{fig:1}
\includegraphics[width=7cm]{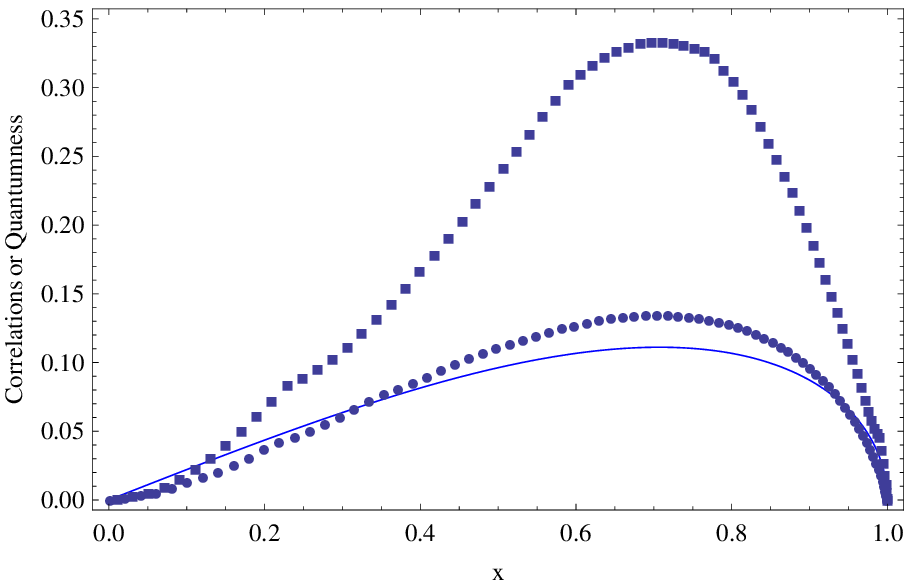}
\caption{The comparison between the quantum correlations and the ensemble
quantumness for a $3\otimes3$ quantum-classical state
$\rho_x=1/3(|0\ra\la0|\otimes|0\ra\la0| +|1\ra \la
1|\otimes|1\ra\la1|+|\phi_2\ra\la
 \phi_2|\otimes|2\ra\la2|)$, where $|\phi_2\ra=\cos\theta|0\ra+\sin\theta|2\ra$ and horizontal
 ordinate is the overlap
 $x=|\la 0|\phi_2\ra|=|\cos\theta|$. Quantum
correlations, quantum discord (circle doted line) and quantum  deficit (square dotted line), and the ensemble quantumness,
non-commutativity (solid line), for $\rho_x$ all get minimal at $x=0,1$ while get
maximal at $x=1/\sqrt{2}$.}
\end{figure}

\medskip
\noindent{\bf Discussions}

We have proposed non-commutativity as a quantumness
measure for an ensemble. It has been shown that the local ensemble
quantumness, instead of quantum correlation measure, like quantum
discord, quantum deficit, etc.,  which is a function of a density
operator, accounts for the local indistinguishability for a complete
set of POPS. It implies that the quantumness of local ensembles must
satisfy certain conditions so that the states in the ensemble cannot be locally
distinguished.

A constructive distinguishing procedure to obtain the concrete local measurements and classical communications  has
been presented to judge the local indistinguishability for a complete set of POPS. Our approach
to judge the local indistinguishability can also be  directly
extended to distinguish a complete set of multipartite POPS or a
non-complete set of POPS, when the local operations are restricted
within nondestructive projective measurements.

Due to that one semi-classical state corresponds to one quantum ensemble, the
non-commutativity has been shown to be able to characterize the
ensemble quantumness for classical-quantum or quantum-classical
systems.

\medskip
\noindent{\bf Methods}

{\sf Proof of the property (2) of the non-commutativity $N$}~Consider two pure $n$
dimensional states $A=|\phi\ra\la \phi|$ and $B=|\psi\ra\la \psi|$.
Denote $\la \phi|\psi \ra=xe^{i\theta}$ with $x\in[0,1]$,
$\theta\in[0,2\pi)$. Then $[A, B]=xe^{i\theta}|\phi\ra \la
\psi|-xe^{-i\theta}|\psi\ra \la \phi|$. Under the base
$\{|\phi\ra=|\phi_0\ra, |\phi_1\ra,...,|\phi_{n-1}\ra\}$, one gets
$|\phi\ra \la \psi|=\sum\limits_{s=0}^{n-1} \la\psi|\phi_s\ra
|\phi_0\ra\la\phi_s|$. $[A, B]$ can then be expressed as
\be
\left(
          \begin{array}{cccc}
            0 & \psi_{0,1} & \ldots & \psi_{0,n-1} \\
            -\psi_{0,1}^*     & 0 & \cdots & 0 \\
            \vdots                &\vdots& \vdots & \vdots \\
            -\psi_{0,n-1}^*   & 0 & \cdots & 0 \\
          \end{array}
        \right),
\ee
where $\psi_{i,j}=\la \phi_i|\psi\ra \la \psi|\phi_j\ra$ and
$\psi_{i,j}^*$ is the conjugate of  $\psi_{i,j}$.  The eigenvalues
of the above matrix are $\{\sqrt{-Z}, -\sqrt{-Z},0,...,0\}$,
 where
 $Z=|\psi_{0,1}|^2+...+|\psi_{0,n-1}|^2=\sum\limits_{j=0}^{n-1}
 |\psi_{0,j}|^2-|\psi_{0,0}|^2=x^2-x^4$. One gets
 $||[A,B]||=2|\sqrt{-Z}|=2x\sqrt{1-x^2}$. Therefore for two pure states $|\psi\ra$ and $|\phi\ra$, $||[|\phi\ra, |\psi\ra]||$
 is 0 when $|\la \phi|\psi\ra|=0$ or $1$, and 1
when $|\la \phi|\psi\ra|={1}/{\sqrt{2}}$.

\medskip
{\sf Proof of Lemma 1}~(a) $\Rightarrow$ (b) First choose any vector in $\varepsilon$, say,
$|a_{i_1}\ra$. Then choose the second vector $|a_{i_2}\ra$ in
$\varepsilon$ such that it is not equal to or orthogonal to
$|a_{i_1}\ra$, i.e., $N(|a_{i_1}\ra, |a_{i_2}\ra)>0$. The existence
of $|a_{i_2}\ra$ is guaranteed by the fact that $\varepsilon$ is a
single set. Next choose the third vector $|a_{i_3}\ra$ in
$\varepsilon$ such that it is independent of $\{|a_{i_1}\ra,
|a_{i_2}\ra\}$, and also not orthogonal to $\{|a_{i_1}\ra,
|a_{i_2}\ra\}$, i.e., $N(|a_{i_1}\ra, |a_{i_2}\ra)<N(|a_{i_1}\ra,
|a_{i_2}\ra, |a_{i_3}\ra)$. Or else, if $|a_{i_3}\ra$ does not
exist, then $\varepsilon$ can be direct sum decomposed into two
parts. Continuing with the above process, we can finally get $m$
linear independent states $\{|a_{i_k}\ra\}_{k=1}^{m}$ in
$\varepsilon$ satisfying inequality (2).

(b) $\Rightarrow$ (c) A nondestructive projective  measurement is
described by an observable, $M$, an Hermitian operator such that all
$|a_i\ra$ in $\varepsilon$ are the eigenvectors of $M$. If
$N(|a_{i_1}\ra, |a_{i_2}\ra)>0$, $|a_{i_1}\ra$ and $|a_{i_2}\ra$ are
neither identical nor orthogonal. Hence the corresponding
eigenvalues of $|a_{i_1}\ra$ and $|a_{i_2}\ra$ must be equal, since
the eigenvectors corresponding to different eigenvalues are
orthogonal. If $N(|a_{i_1}\ra, |a_{i_2}\ra)<N(|a_{i_1}\ra,
|a_{i_2}\ra,|a_{i_3}\ra)$, $|a_{i_3}\ra$ is at least not orthogonal
to one of $|a_{i_1}\ra$ and $|a_{i_2}\ra$. For instance, suppose
$N(|a_{i_1}\ra,|a_{i_3}\ra)>0$, then the eigenvalues corresponding
to $|a_{i_1}\ra$ and $|a_{i_3}\ra$ are equal. Therefore we have that
if (2) is satisfied, those $m$ linear independent $|a_{i_k}\ra$s
have the same eigenvalues of $M$. Therefore $M$ becomes an identity
(apart of real factor) and it cannot be used to distinguish the
states in $\varepsilon$.

(c) $\Rightarrow$ (a) Suppose the set is not a single one and has a
decomposition $\varepsilon=\varepsilon_1 \oplus \varepsilon_2$. Then
one can use a  corresponding nondestructive projective measurement
$M=aI_1\oplus bI_2$, where $a\neq b$, $I_1, I_2$ are identity
operators, to distinguish the states in $\varepsilon_1$ from
$\varepsilon_2$. Hence if $M$ cannot do any in distinguishing the
states in $\varepsilon$, $\varepsilon$ must be a single set, that
is, statement (a) holds.

\medskip
{\sf The unique expression of semi-classical states}~Let $\rho=\sum\limits_i p_i\rho_i\otimes|\alpha_i\ra \la \alpha_i|$ be
an arbitrary given semi-classical state, where $\{|\alpha_i\ra\}$ is
an orthogonal base. Suppose there exists another orthogonal base
$\{|\beta_i\ra\}$ such that $\rho=\sigma$, where
$\sigma=\sum\limits_i q_i\sigma_i\otimes|\beta_i\ra \la\beta_i|$.
Let $U=\sum\limits_i|\beta_i\ra \la \alpha_i|$ be the unitary
operator such that $U|\alpha_i\ra=|\beta_i\ra$, $(U)_{ij}=u_{ij}=\la
\alpha_i|\beta_j \ra$, $|\beta_i\ra=\sum\limits_s
u_{si}|\alpha_s\ra$. So $\sigma$ can be reexpressed as
$\sum\limits_{st}(\sum\limits_i q_i \sigma_i u_{si} u_{ti}^*)
\otimes |\alpha_s\ra \la \alpha_t|$. Let $\widetilde{q}^i_{kl}$ be
the entries of $q_i\sigma_i$. Set $Q_{kl}=\sum\limits_i
\widetilde{q}^i_{kl} |\beta_i\ra\la \beta_i|$. Then $\rho=\sigma$
implies that
\be
\begin{aligned}
0=&\sum\limits_i \widetilde{q}^i_{kl} u_ {si} u_{ti}^*=\sum\limits_i \widetilde{q}^i_{kl} \la \alpha_s|\beta_i\ra {\la
\alpha_t|\beta_i\ra}^*
\\=&\la \alpha_s|(\sum\limits_i
\widetilde{q}^i_{kl}|\beta_i\ra\la \beta_i|)|\alpha_t\ra
=\la\alpha_s|Q_{kl}|\alpha_t\ra,
\end{aligned}
\ee
for $s\neq t$, which means that both $\{|\alpha_i\ra\}$ and
$\{|\beta_i\ra\}$ are the eigenvectors of $Q_{kl}$, and
$\widetilde{q}^i_{kl}$ are the corresponding eigenvalues. Note that
the set  $\{q_i\sigma_i\}$ of state $\sigma$ can be divided into
degenerate part (in which all the states are identical) and
non-degenerate part (in which all the states are mutually
different). For non-degenerate part, taking over all $k,l$, one will
find that the intersection of the eigenspaces belonging to the
eigenvalues $\widetilde{q}^i_{kl}$ must be one dimensional, namely,
$|\alpha_i\ra=|\beta_i\ra$ and then $p_i\rho_i=q_i\sigma_i$. For
degenerate part, without loss of generality, assume
$q_1\sigma_1=q_2\sigma_2$, then $\rho=\sigma$ means
$p_1\rho_1\otimes|\alpha_1\ra \la
\alpha_1|+p_2\rho_2\otimes|\alpha_2\ra \la \alpha_2|
=q_1\sigma_1\otimes(|\beta_1\ra \la \beta_1|+|\beta_2\ra \la
\beta_2|)$. Therefore $p_1\rho_1=p_2\rho_2=q_1\sigma_1=q_2\sigma_2$.
Finally we get $p_i\rho_i=q_i\sigma_i, \forall{i}$.

\newpage
\bigskip
\noindent{\sf Acknowledgements}

\noindent The work is supported by NSFC under number 11275131. The
authors thank X.N. Zhu,  Z.X. Wang, and W.P. Yang for useful discussions.

\bigskip
\noindent{\sf Author contributions}

\noindent  T.M. and M.Z. and Y.W. and S.M. wrote the main manuscript text. All authors reviewed the manuscript.

\bigskip
\noindent{\sf Additional Information}

\noindent Competing Financial Interests: The authors declare no competing financial interests.

\end{document}